\documentclass{vgtc}

\ifpdf
  \pdfoutput=1\relax                   
  \usepackage{graphicx}                
  \DeclareGraphicsExtensions{.pdf,.png,.jpg,.jpeg} 
\else
  \usepackage{graphicx}                
  \DeclareGraphicsExtensions{.eps}     
\fi%

\graphicspath{{figures/}{pictures/}{images/}{./}} 

\usepackage{microtype}                 
\PassOptionsToPackage{warn}{textcomp}  
\usepackage{textcomp}                  
\usepackage{mathptmx}                  
\usepackage{times}                     
\usepackage{cite}                      
\usepackage{tabu}                      
\usepackage{booktabs}                  
\usepackage{amsmath}
\usepackage[nolist]{acronym}
\usepackage{amsmath}
\usepackage{listings}
\usepackage[table,xcdraw]{xcolor}
\usepackage{xcolor}
\definecolor{maroon}{cmyk}{0, 0.87, 0.68, 0.32}
\definecolor{halfgray}{gray}{0.55}
\definecolor{ipython_frame}{RGB}{207, 207, 207}
\definecolor{ipython_bg}{RGB}{247, 247, 247}
\definecolor{ipython_red}{RGB}{186, 33, 33}
\definecolor{ipython_green}{RGB}{0, 128, 0}
\definecolor{ipython_cyan}{RGB}{64, 128, 128}
\definecolor{ipython_purple}{RGB}{170, 34, 255}

\usepackage{listings}
\lstdefinelanguage{iPython}{
    morekeywords={access,and,break,class,continue,def,del,elif,else,except,exec,finally,for,from,global,if,import,in,is,lambda,not,or,pass,print,raise,return,try,while},%
    morekeywords=[2]{abs,all,any,basestring,bin,bool,bytearray,callable,chr,classmethod,cmp,compile,complex,delattr,dict,dir,divmod,enumerate,eval,execfile,file,filter,float,format,frozenset,getattr,globals,hasattr,hash,help,hex,id,input,int,isinstance,issubclass,iter,len,list,locals,long,map,max,memoryview,min,next,object,oct,open,ord,pow,property,range,raw_input,reduce,reload,repr,reversed,round,set,setattr,slice,sorted,staticmethod,str,sum,super,tuple,type,unichr,unicode,vars,xrange,zip,apply,buffer,coerce,intern},%
    sensitive=true,%
    morecomment=[l]\#,%
    morestring=[b]',%
    morestring=[b]",%
    morestring=[s]{'''}{'''},
    morestring=[s]{"""}{"""},
    morestring=[s]{r'}{'},
    morestring=[s]{r"}{"},%
    morestring=[s]{r'''}{'''},%
    morestring=[s]{r"""}{"""},%
    morestring=[s]{u'}{'},
    morestring=[s]{u"}{"},%
    morestring=[s]{u'''}{'''},%
    morestring=[s]{u"""}{"""},%
    literate=
    {á}{{\'a}}1 {é}{{\'e}}1 {í}{{\'i}}1 {ó}{{\'o}}1 {ú}{{\'u}}1
    {Á}{{\'A}}1 {É}{{\'E}}1 {Í}{{\'I}}1 {Ó}{{\'O}}1 {Ú}{{\'U}}1
    {à}{{\`a}}1 {è}{{\`e}}1 {ì}{{\`i}}1 {ò}{{\`o}}1 {ù}{{\`u}}1
    {À}{{\`A}}1 {È}{{\'E}}1 {Ì}{{\`I}}1 {Ò}{{\`O}}1 {Ù}{{\`U}}1
    {ä}{{\"a}}1 {ë}{{\"e}}1 {ï}{{\"i}}1 {ö}{{\"o}}1 {ü}{{\"u}}1
    {Ä}{{\"A}}1 {Ë}{{\"E}}1 {Ï}{{\"I}}1 {Ö}{{\"O}}1 {Ü}{{\"U}}1
    {â}{{\^a}}1 {ê}{{\^e}}1 {î}{{\^i}}1 {ô}{{\^o}}1 {û}{{\^u}}1
    {Â}{{\^A}}1 {Ê}{{\^E}}1 {Î}{{\^I}}1 {Ô}{{\^O}}1 {Û}{{\^U}}1
    {œ}{{\oe}}1 {Œ}{{\OE}}1 {æ}{{\ae}}1 {Æ}{{\AE}}1 {ß}{{\ss}}1
    {ç}{{\c c}}1 {Ç}{{\c C}}1 {ø}{{\o}}1 {å}{{\r a}}1 {Å}{{\r A}}1
    {€}{{\EUR}}1 {£}{{\pounds}}1,
    literate=
    *{+}{{{\color{ipython_purple}+}}}1
    {-}{{{\color{ipython_purple}-}}}1
    {*}{{{\color{ipython_purple}$^\ast$}}}1
    {/}{{{\color{ipython_purple}/}}}1
    {^}{{{\color{ipython_purple}\^{}}}}1
    {?}{{{\color{ipython_purple}?}}}1
    {!}{{{\color{ipython_purple}!}}}1
    {\%}{{{\color{ipython_purple}\%}}}1
    {<}{{{\color{ipython_purple}<}}}1
    {>}{{{\color{ipython_purple}>}}}1
    {|}{{{\color{ipython_purple}|}}}1
    {\&}{{{\color{ipython_purple}\&}}}1
    {~}{{{\color{ipython_purple}~}}}1
    {==}{{{\color{ipython_purple}==}}}2
    {<=}{{{\color{ipython_purple}<=}}}2
    {>=}{{{\color{ipython_purple}>=}}}2
    {+=}{{{+=}}}2
    {-=}{{{-=}}}2
    {*=}{{{$^\ast$=}}}2
    {/=}{{{/=}}}2,
    commentstyle=\color{ipython_cyan}\ttfamily,
    stringstyle=\color{ipython_red}\ttfamily,
    keepspaces=true,
    showspaces=false,
    showstringspaces=false,
    rulecolor=\color{ipython_frame},
    frame=single,
    frameround={t}{t}{t}{t},
    framexleftmargin=0mm,
    numbers=left,
    numberstyle=\tiny\color{halfgray},
    backgroundcolor=\color{ipython_bg},
    basicstyle=\scriptsize\ttfamily,
    keywordstyle=\color{ipython_green}\ttfamily,
    escapechar=\¢,escapebegin=\color{ipython_green},
}

\usepackage{flushend}

\usepackage[square,sort,comma,numbers]{natbib}
\usepackage{soul}
\usepackage{amsmath}
\usepackage{amssymb}
\usepackage{wrapfig}
\usepackage{booktabs}
\usepackage{multirow}
\usepackage{xspace}

\def\figurePath{images/}

\def\mycfigure#1#2{%
    \begin{figure*}[htb]%
    \centering\includegraphics*[width = \linewidth]{\figurePath#1}%
    \vspace{-.2cm}%
    \caption{#2}%
    \label{fig:#1}%
    \end{figure*}%
}

\newcommand{\eg}{e.\,g.,\ }
\newcommand{\ie}{i.\,e.,\ }
\newcommand{\etal}{et~al.\ }

\newcommand{\refSec}[1]{Sec.~\ref{sec:#1}}
\newcommand{\refFig}[1]{Fig.~\ref{fig:#1}}

\newcommand{\refLst}[1]{Listing~\ref{lst:#1}}

\soulregister\ref7
\soulregister\cite7
\soulregister\refFig7
\soulregister\refAlg7
\soulregister\cite7
\soulregister\ref7
\soulregister\pageref7
\soulregister\shortcite7
\soulregister\eg0
\soulregister\ie0
\soulregister\etal0

\DeclareGraphicsExtensions{.png,.jpg,.pdf,.ai,.psd}
\newcommand{\mysection}[2]{\section{#1}\label{sec:#2}}
\newcommand{\mysubsection}[2]{\subsection{#1}\label{sec:#2}}

\newcommand{\mymath}[2]{\newcommand{#1}{\TextOrMath{$#2$\xspace}{#2}}}
\usepackage{duckuments}             

\onlineid{1536}

\vgtccategory{Technology}
\vgtcinsertpkg

\title{Metameric Varifocal Holograms}

\newif\ifarxiv
\arxivfalse

\newif\ifanon
\anonfalse

\author{David R.~Walton\thanks{e-mail: david.walton.13@ucl.ac.uk}\\ %
        \scriptsize UCL %
\and Koray Kavaklı\thanks{e-mail: kkavakli@ku.edu.tr}\\ %
     \scriptsize Koç University
\and Rafael Kuffner dos Anjos\thanks{e-mail: r.kuffnerdosanjos@leeds.ac.uk}\\ %
     \scriptsize University of Leeds
\and David Swapp\thanks{e-mail: d.swapp@ucl.ac.uk}\\ %
     \scriptsize UCL
\and Tim Weyrich\thanks{e-mail: t.weyrich@cs.ucl.ac.uk}\\ %
     \scriptsize UCL
\and Hakan Urey\thanks{e-mail: hurey@ku.edu.tr}\\
     \scriptsize Koç University
\and Anthony Steed\thanks{e-mail: a.steed@ucl.ac.uk}\\
     \scriptsize UCL
\and Tobias Ritschel\thanks{e-mail: t.ritschel@ucl.ac.uk}\\
     \scriptsize UCL
\and Kaan Ak\c{s}it\thanks{e-mail: k.aksit@ucl.ac.uk}\\
     \scriptsize UCL
     }

\ifarxiv
\renewcommand{\manuscriptnotetxt}{}
\fi

\abstract{
Computer-Generated Holography (CGH) offers the potential for genuine, high-quality three-dimensional visuals.
However, fulfilling this potential remains a practical challenge due to computational complexity and visual quality issues.
We propose a new CGH method that exploits gaze-contingency and perceptual graphics to accelerate the development of practical holographic display systems.
Firstly, our method infers the user's focal depth and generates images only at their focus plane without using any moving parts.
Second, the images displayed are metamers; in the user's peripheral vision, they need only be statistically correct and blend with the fovea seamlessly.
Unlike previous methods, our method prioritises and improves foveal visual quality without causing perceptually visible distortions at the periphery.
To enable our method, we introduce a novel metameric loss function that robustly compares the statistics of two given images for a known gaze location.
In parallel, we implement a model representing the relation between holograms and their image reconstructions.
We couple our differentiable loss function and model to  metameric varifocal holograms using a stochastic gradient descent solver.
We evaluate our method with an actual proof-of-concept holographic display, and we show that our CGH method leads to practical and perceptually three-dimensional image reconstructions.
} 


\CCScatlist{
\CCScatTwelve{Computing methodologies}{Computer graphics}{Graphics systems and interfaces}{Perception};
\CCScatTwelve{Hardware}{Communication hardware, interfaces and storage}{Displays and imagers}{}
}

\teaser{
  \centering
  \includegraphics[width=\linewidth]{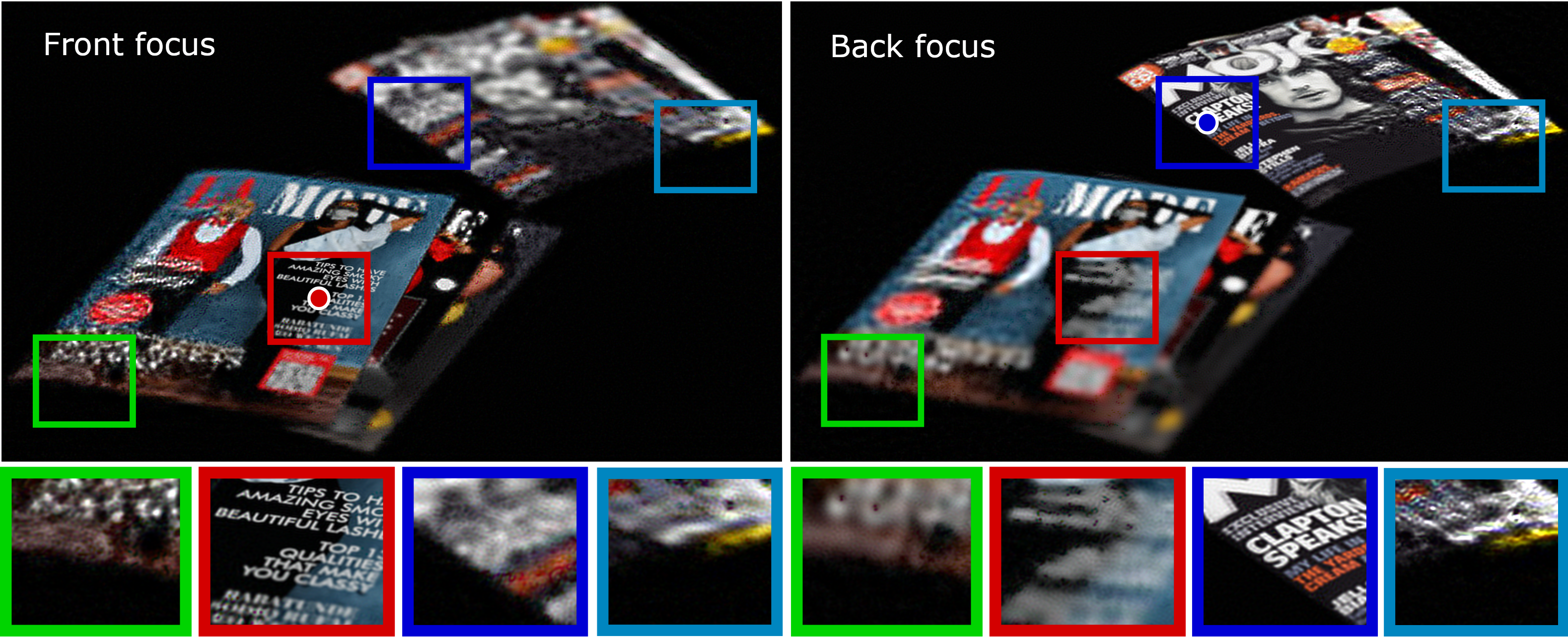}
  \caption{
  Simulated reconstuctions of metameric varifocal holograms.
  Our holograms reconstruct single-plane images at the correct focus levels, reconstructing high-resolution visuals at a user's fovea while displaying statistically correct content across their peripheral vision indistinguishable from the target images (metamers). 
  Top row: simulated image reconstructions at two different focus levels (gaze location marked with a dot). 
  Bottom row: zoomed-in insets from these two reconstructions. 
  All foveated images in this paper are best viewed at a 60 cm wide display from a distance of 80 cm.
  (Three-dimensional assets from Vilém Duha \textcopyright 2021)
  } 
  \label{fig:teaser}
}

\mymath{\image}{I}
\mymath{\analysis}{A}
\mymath{\synthesis}{S}
\mymath{\stats}{s}
\mymath{\gaze}{\mathbf g}
\mymath{\pixelsize}{p}
\mymath{\metamerloss}{\mathcal{M}}
\mymath{\fouriertransform}{\mathcal{F}}
\mymath{\amplitude}{A}
\mymath{\phase}{\phi}
\mymath{\wave}{u}
\mymath{\wavelength}{\lambda}
\mymath{\distance}{z}
\mymath{\propagator}{\mathsf H}
\mymath{\perception}{\mathsf P}
\mymath{\blurlow}{\mathcal{B}_l}
\mymath{\blurmatch}{\mathcal{B}_m}
\mymath{\metamerl}{\mathcal{M}_{L2}}
\newcommand{\perceived}{_\mathrm{p}}
\newcommand{\target}{_\mathrm{t}}

\begin{document}

\begin{acronym}
\acro{AdaIN}{Adaptive Instance Normalisation}
\acro{CGH}{Computer-Generated Holography}
\acro{HCI}{Human-Computer Interaction}
\acro{HVS}{Human Visual System}
\acro{LoD}{Level of Detail}
\acro{MSE}{Mean Squared Error}
\acro{SLM}{Spatial Light Modulator}
\acro{CFF}{Critical Flicker Fusion}
\acro{DoF}{Depth of Field}
\acro{GI}{Global Illumination}
\acro{VGG}{Visual Geometry Group}
\end{acronym}


\firstsection{Introduction}

\maketitle

In recent years, improving display technology to enable lifelike three-dimensional visuals has attracted much attention from industry and academia as displays are crucial for future \ac{HCI}~\cite{orlosky2021telelife}.
An emerging trend, \ac{CGH}~\cite{slinger2005computer}, promises such realistic visuals in the next-generation of displays~\cite{koulieris2019near}.
Unlike conventional displays, pixelated images are not sent to holographic displays directly. 

In a typical phase-only \ac{SLM}-based holographic display, laser light illuminates an array of pixels which modulate the phase of the light. 
The reflected light interferes to produce the image. 
Finding the correct phase values to send to the \ac{SLM} is challenging due to the complexity of light transport.
Also, \ac{SLM}s have limited resolution.
As a result, real \ac{CGH} displays suffer from noise and other artefacts.


Gaze-contingent approaches~\cite{kim2019foveated,aksit2019manufacturing} are often used to reduce the hardware and computational requirements of displays.
In this work, we explore whether gaze-contingency for \ac{CGH} can help meet the demands of the \ac{HVS} in practice.
Knowing the user's gaze gives us two critical pieces of information we exploit.

First, it tells us which parts of the image fall in the user's periphery, rather than their fovea.
To exploit this, we draw inspiration from the state of the art in foveated graphics literature~\cite{walton2021beyond}.
This work focuses on generating visuals which are not pixel-accurate to a target image in the periphery of the user's vision, but are still perceived as identical to the target.
We exploit their work to dedicate more of the expressive power of the \ac{SLM} to generating high-quality visuals at the fovea as described in work by Chakravarthula et al.~\cite{chakravarthula2021gaze}.
In contrast, visuals at the periphery need only be statistically correct (in a sense precisely described in \refSec{PerceptionModel}) and will still be perceived as accurate.
As highly accurate simulation models become available in the future, such a method can pave the way towards distributing the speckle noise at a holographic display~\cite{curtis2021dcgh} in a statistically correct way, enabling indistinguishable images at the periphery in the future.

Second, given the depths of each pixel in the displayed image, it allows us to infer the user's current focal depth.
We can use this information to only enforce our reconstruction to be correct at the user's current focus.
Whilst \ac{CGH} is certainly capable of displaying multi-plane images, this often leads to image quality issues as the hologram pixels are used to deliver images at multiple planes at once.
For that purpose, we draw inspiration from existing literature on varifocal near-eye displays~\cite{aksit2017near,padmanaban2017optimizing} and varifocal holograms~\cite{peng2020neural}. 
We argue that generating images at a single plane instead of multiple planes will help assure quality in visuals generated by \ac{CGH}. We combine these arguments to enable \ac{CGH} computation pipelines that are perceptually accurate and offer high visual quality.

Specifically, this work introduces the following contributions: 

\noindent\textbf{(1) Metameric loss function.} We introduce a fast metameric loss that can help us quantify image quality within the peripheral field of view by comparing the statistics of images. 
We believe this loss couples well with a gaze-contingent display and graphics application, specifically holographic displays, as they are often proposed as the next-generation display technology.

\noindent\textbf{(2) Metameric varifocal holograms.} We introduce a complete optimisation pipeline for metameric varifocal holograms using our metameric loss function. 
Note that our holograms change focus in a gaze-contingent manner, avoiding the complexity of representing light fields or multiplane images using \ac{CGH};

\noindent\textbf{(3) Proof-of-concept prototype.}
We build a single colour holographic display to experiment with our metameric varifocal holograms.
We assess the results of our \ac{CGH} method using this proof-of-concept display.

\section{Related Work}
Our work combines the state of the art in visual perception and \ac{CGH} while relying on gaze contingency.
Hence, we review the relevant work in visual perception, gaze-contingent displays and \ac{CGH} fields.

\subsection{Gaze-contingent displays}
Eye-gaze tracking~\cite{li2020optical,angelopoulos2020event} is of great interest to AR and VR research.
A major reason for this is that visual~\cite{elliott1995visual} and depth acuity~\cite{wang2004depth} of the \ac{HVS} drops sharply with increasing eccentricity towards far peripheral vision.
Combined with the visual and depth acuity of \ac{HVS}, eye-gaze information opens up opportunities towards reducing computational and hardware complexity of displays.
Such displays that take advantage of eye-gaze information are known as gaze-contingent displays.

A form of gaze-contingent displays -- foveated displays~\cite{spjut2020toward} -- present images at high resolution at the fovea and lower resolution at the periphery.
A foveated display tracks a user's gaze and can either actively move a foveal inset display~\cite{tan2018foveated}, move both foveal and peripheral insets~\cite{kim2019foveated}, or change the distribution of resolution by distorting optical fields~\cite{aksit2019manufacturing} to generate images with fewer pixels but with no perceptual difference.

Our work falls into the category of foveated displays.
It merges the ideas of distorting~\cite{aksit2019manufacturing} or shifting~\cite{aksit2017near} optical fields while taking advantage of \ac{CGH} in a foveated manner following the spirit of the work by Chakravarthula et al.~\cite{chakravarthula2021gaze}.
A major advantage of \ac{CGH} in this setting is that it can facilitate foveated rendering without the need for moving parts in the form of displays or lenses.

\subsection{Metamers}

Most popular foveated rendering approaches focus on decreasing resolution with increasing eccentricity~\cite{friston2019perceptual, wang2021focas}.
However, traditional literature on human vision~\cite{aubert1857beitrage} refers to the objects in the periphery as difficult to see and different, but not particularly blurry. 
Objects are not only less sharp, but the size of stimuli~\cite{ziemba2021opposing}, visual crowding~\cite{gong2018extraction} and texture content~\cite{wallis2018image} also play an important part in how things are perceived. 
For a comprehensive review of the behavior and the limitations of peripheral vision, we recommend the review from Rozenholtz~\cite{rosenholtz2016capabilities}.

\begin{figure}
\centering
\includegraphics[width=\columnwidth]{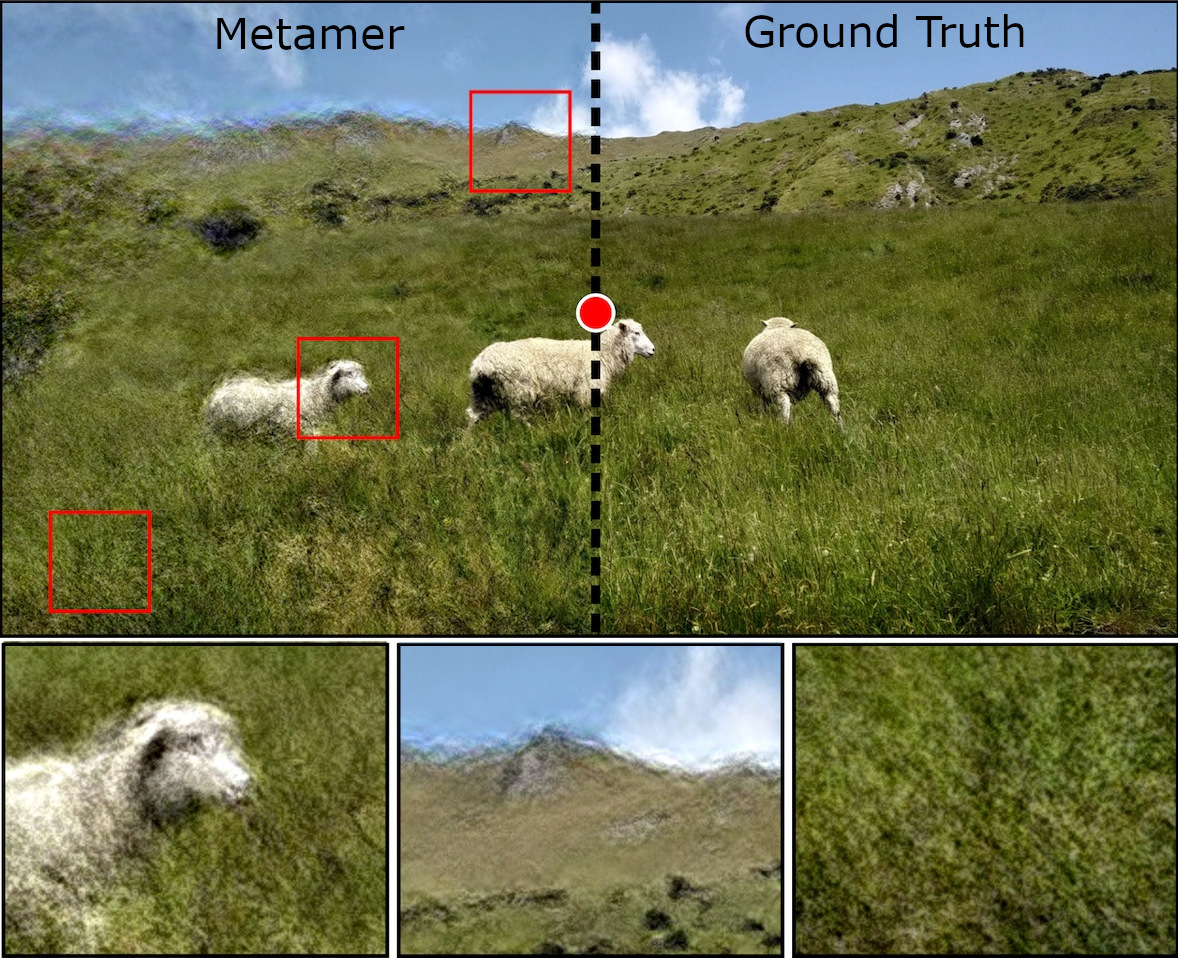}
\caption{
A sample metamer display at the top row is generated by following the work by Walton et al.~\cite{walton2021beyond} using a gaze location at the center of the image (red dot at the center). 
For comparison purposes, we also show the ground truth image at the right portion of the same image.
Red highlighted regions from the top image are zoomed-in and provided as insets at the bottom row.
}
\label{fig:metamer-example}
\end{figure}

With this in mind, Freeman and Simoncelli~\cite{freeman2011metamers} showed that it is possible to devise a process to generate ``ventral metamers''; pairs of images perceived as identical for a given fixation point (see \refFig{metamer-example}).
Their work~\cite{freeman2011metamers} models the correlation between the size of pooling regions and different eccentricities, describing visuals in the periphery with local image statistics rather than individual pixel values.
Unfortunately, their process is computationally expensive as it depends on complex image statistics and iterative optimisation processes.
Deza et al.~\cite{deza2019towards} proposed an approach to approximate this effect using techniques from style transfer~\cite{gatys2016image}, blending \ac{VGG} network~\cite{simonyan2014very} features of the target image with those from a noise image using \ac{AdaIN}\cite{huang2017arbitrary} over foveated pooling regions. 
Similarly, recent work from Surace et al.~\cite{surace2021learning} uses a texture synthesis approach combined with Generative Adversarial Networks (GANs) to generate ventral metamers.
There is no guarantee that the generated visuals will be statistically correct for any known human vision model in both cases.
These techniques are also unable to operate at interactive framerates, although they are significantly faster than that of Freeman and Simoncelli~\cite{freeman2011metamers}.

To our knowledge, the first work that achieves generation of ventral metamers at interactive rates is the work by Walton et al.~\cite{walton2021beyond}, which uses a simplified statistical model focused on using fast calculated means and variances of a steerable pyramid~\cite{portilla2000parametric}.
Their simplified model allows fast synthesis of metamers by scaling and biasing bands of a steerable pyramid constructed from a noise image.

Our work reformulates their model~\cite{walton2021beyond} as a general-purpose differentiable loss function.
This opens it up to a range of other applications, including but not limited to \ac{CGH} as described in this paper.

\subsection{Computer-Generated Holography}
\ac{CGH} has garnered much interest from the research community in recent years.
This interest primarily stems from the widespread availability of powerful, highly parallel processors coupled with modern machine-learning frameworks that automatically differentiate given models~\cite{paszke2017automatic}.
These advancements accelerate and improve the accuracy of hologram generation (phase retrieval), particularly when taking advantage of advances in deep learning~\cite{zhang2021phasegan}.
As a result, modern phase retrieval techniques offer dramatically improved image quality over classical hologram calculation methods such as the Gerchberg-Saxton method~\cite{yang1994gerchberg}.
From the recent past, the work by Chakravarthula et al.~\cite{chakravarthula2019wirtinger} revisits Wirtinger complex derivatives and shows that the visual quality of two-dimensional image reconstructions in \ac{CGH} can be improved in common Gerchberg-Saxton and Double-phase coding~\cite{hsueh1978computer} approaches.
The works by Peng et al.~\cite{peng2020neural} and Chakravarthula et al.~\cite{chakravarthula2020learned} help to bridge the gap between \ac{CGH} simulations and actual image reconstructions in a physical display by learning a model of display hardware using a camera and convolutional neural networks.
Their findings have drastically improved the quality of two-dimensional image reconstructions in actual holographic displays.
Closest to our work, Chakravarthula et al.~\cite{chakravarthula2021gaze} show that na\"ive foveation using an importance map can help to combat speckle in two-dimensional \ac{CGH} image reconstruction.
Their technique does not guarantee peripheral visuals that are indistinguishable from a target image, however.

Meanwhile, three-dimensional image reconstructions in holographic displays have also seen dramatic improvement.
The work by Maimone et al.~\cite{maimone2017holographic} represents each point in a three-dimensional scene by adding a relevant sub-hologram to a final hologram.
Their work exploits separable functions and introduces superior image quality with their fast pointwise method.
Meanwhile, alternative approaches~\cite{shi2017near} that treat a three-dimensional scene as a multi-view image stack also prove themselves in terms of image quality.
However, all of these approaches are computationally expensive, do not yet run at interactive rates and require a high level of sophistication in data representations.
The work by Shi et al.~\cite{shi2021towards} shows that pointwise approaches can potentially run at interactive rates with a learning methodology that cleverly stitches occluded sub-holograms to a final hologram.
Thus, their work paves the way towards three-dimensional \ac{CGH} at interactive rates in the future.

Our work deals with two-dimensional image reconstructions in \ac{CGH}.
However, unlike existing work, our holograms actively change the depth plane with a user's focus and use state-of-the-art perceptual graphics to maintain the highest visual quality possible.
In addition, our work does not require sophisticated data representations (we operate on images rather than three-dimensional data).
To our knowledge this is a unique combination, and we believe it can pave a path to practical \ac{CGH}.

\mycfigure{Setup}{
Overview of our system comprised of a display model \emph{(left)} and a perception model \emph{(right)}.
This scheme shows a specific fixation \gaze and hence a specific propagation $\propagator(\distance)$ for a specific focus \distance out of many possible changes over time as we gaze and focus.\vspace*{-1ex}}

\mysection{Method}{Method}
Our simulation pipeline is composed of two primary blocks, a holographic display model (\refSec{Display}) and a perceptual model (\refSec{PerceptionModel}) which is used to define our metameric loss.
Both blocks are differentiable and have no tunable or learned components.
Our work aims to optimise the input to the display (phase values, \phase) such that the difference between the resulting percept and the percept of a reference image $\image\target$ is minimised (\refFig{Setup}).
To this end, we rely on a display model $\propagator$ and a model of perception $\perception$.

The display model $\propagator(\distance)$ will map phase values to image intensities at a certain distance \distance.
Note that in a varifocal display~\cite{aksit2017near,dunn2017wide,rathinavel2019varifocal}, the required focus \distance is assumed to be known at every frame.
How light is propagated will depend on that focus distance.

The perception model \perception maps image intensities into a perceptual space where distances between points are perceptually uniform~\cite{freeman2011metamers,walton2021beyond}.
This means that images which are perceived as similar should map to nearby points in the perceptual space, and images perceived as dissimilar should map to more distant points.

Putting both display and perception model together, we optimise
\begin{equation}\label{eqn:optimise}
\mathrm{argmin}_{\phase}
\metamerloss(\propagator(\distance)\cdot\phase, 
\image\target)
.
\end{equation}

where $\metamerloss$ is our metameric loss function, defined as:
\begin{equation}\label{eqn:metameric_loss}
\metamerloss(A, B) :=
|
\perception(A)-
\perception(B)
|_2
.
\end{equation}

We provide details of our display and perceptual model in the following \refSec{Display} and \refSec{PerceptionModel}.

\mysubsection{Display system and model}{Display}
We use a combination of an actual holographic display and a commonly accepted differentiable model of that display hardware.

\noindent\textbf{System:}We design phase-only diffractive components that can either be represented with a static diffractive optical element~\cite{swanson1989binary} or a programmable phase-only \ac{SLM}~\cite{savage2009digital}.
While amplitude modulation physically blocks light, phase-only holograms are a light-efficient form of optical beam shaping.
More sophisticated versions of these components such as cascaded~\cite{mengu2019analysis} or volumetric~\cite{jang2020design} holograms do not fall into the scope of this work.
Importantly, such a system can be reliably modeled using a differentiable operator, explained next.

\noindent\textbf{Model:} The relation between a complex light wave with unit amplitude and phase \phase leaving our phase-only \ac{SLM} and the light wave $\wave(\distance)$ at the image plane depth $\distance$ is described by the Fresnel diffraction operator $\propagator$ as $\wave(\distance)=\propagator(\distance) \cdot \phase$.
For derivation of this operator see~\cite{born2013principles} and section 1 of the supplementary material for implementation details.
Notably, $\mathsf H$ is a linear operator and hence differentiable and can be used to optimise holograms~\cite{zhang20173d}.

In our case, \distance varies when the user changes gaze \gaze but is fixed for any point in time where there is a unique gaze and a unique focus.
The \ac{HVS} perceives intensity (wave amplitude-square) of light; therefore, the perceived reconstructed image, $\image\perceived$ is
\begin{equation}
\image\perceived=|\propagator(\distance) \cdot \phase|^2.
\end{equation}
A differentiable PyTorch implementation of this model is readily available \ifanon at REMOVEDFORANONYMITY \else in the odak library \cite{kaan_aksit_2021_5136406} \fi.
Using this library and work by \ifanon REMOVEDFORANONYMITY \else Kavakl{\i} et al.~\cite{kavakli2021learned}\fi, it is also possible to optimise this $\mathsf H$ operator in a camera-in-the-loop fashion to produce the best possible output on an actual holographic display.
In our pipeline, we use an $\mathsf H$ operator optimised in this way to best suit our display hardware.

\mysubsection{Perception model}{PerceptionModel}

Our perceptual model maps images to a perceptual space as outlined above.
This mapping forms the core of our metameric loss, which is computed by measuring the distance between two images after transforming both to this perceptual space.
In this work, we strive to make our model efficient (following \cite{walton2021beyond}) and also differentiable, allowing it to be used effectively for any optimisation or machine learning task requiring foveated output.

Our perceptual model is inspired by the analysis step of \cite{walton2021beyond}, with some alterations to make it more suitable for backpropagation (see supplementary material for further details). 
When processing colour images, as in \cite{walton2021beyond} these are first converted to a YCbCr colourspace~\cite{poynton2012digital}.
We then compute the (real-valued) steerable pyramid~\cite{portilla2000parametric} of an input image. 
From each level, $i$ of this pyramid, local statistics $s_i$ are then computed.
These $s_i$ consist of means and variances computed over local pooling regions around each pixel which grow with eccentricity.
We first describe how these local statistics are computed, then describe how the size of the pooling regions is determined.

Local means of an image can be found by convolving the image with a normalised low-pass filter $F$.
To determine local variances, we use the identity:
\begin{equation}
\mathbb V[X] = \mathbb E[X^2] - \mathbb E[X]^2
\end{equation}

Thus, we identify the local variances by applying another lowpass filter $F$, squaring and subtracting: $\mathbb V[\image] = F\ast(\image^2) - (F\ast(\image))^2$.

The bandwidth of the lowpass filter decreases with increasing eccentricity.
We set the angular pooling size to be proportional to the square of the eccentricity, as we found this gave the best results with the method of \cite{walton2021beyond}.
The constant of proportionality $\alpha$ is the parameter of the approach that controls the foveation effect's aggressiveness. 
In practice, to accelerate this spatially-varying lowpass filter, as in \cite{walton2021beyond}, we compute a MIP map of the input image that we sample using trilinear interpolation to achieve the correct pooling size at each pixel.

At each pixel, we determine the \ac{LoD} value to use in this sampling so the pooling region covers the appropriate angular size. 
Full details of this specific process are provided in the supplementary material.

Since pooling size is proportional to eccentricity squared, some pixels near the fixation point will have pooling sizes less than or equal to one pixel.
For these pixels, rather than measuring loss over all pyramid levels we calculate a direct $\mathcal{L}_2$ loss against the target image.
In principle this does not change the loss function, but in practice we found it helped the most critical part of the image near the fovea to converge to a noise-free result more quickly.

Our loss function is differentiable and accelerated through a GPU implementation using a modern machine learning library with automatic differentiation~\cite{paszke2017automatic}. 
It is straightforward to use it directly in any desired optimisation/training (a PyTorch implementation is available at \ifanon REMOVEDFORANONYMITY\else \cite{kaan_aksit_2021_5136406}\fi).

\section{Implementation}\label{sec:Implementation}
The implementation of our proposal contains two building blocks.
These are the actual hologram optimisation pipeline and a proof-of-concept holographic display prototype.
We provide details of each in the next sections.

\subsection{Hologram optimisation pipeline}
We implement a differentiable metameric varifocal hologram optimisation pipeline using a modern machine learning library with automatic differentiation~\cite{paszke2017automatic}.
Source code of our implementation is publicly available at \ifanon REMOVEDFORANONYMITY \else \cite{metaholo_repo} \fi.

\noindent\textbf{Target Image Capture/Generation:}
We use both real captured images and virtual rendered images as target images for our optimisation. 
In both cases, we intentionally limit the \ac{DoF} of the images to mimic the optics of the human eye.
We note that images formed on the retina already contain \ac{DoF} blur due to the optics of the eye. 
Metamerisation relies upon the processing in the \ac{HVS} that takes place right after the optics of the eye.
We mimic this approach by applying a metameric loss to a target with limited \ac{DoF}, to replicate the physical process more accurately. 
If the target images were sharp everywhere, our output reconstructions would be sharp in regions of the image which would have \ac{DoF} blur if the real scene were viewed by the user, giving unrealistic output.

When capturing real images, we adjust the depth of field of our camera by opening the aperture to the degree that best approximates the \ac{DoF} we observed when viewing the real scene (\textit{\textflorin}-number 2 to 8).
Virtual scenes were rendered using the Cycles raytracer in Blender \cite{blender2021blender}, where we can render realistic scenes with \ac{GI}, and also simulating \ac{DoF} blur (rendering scripts are available at \ifanon REMOVEDFORANONYMITY\else \cite{metaholo_repo})\fi.


\begin{minipage}{.45\textwidth}
\begin{lstlisting}[language=iPython,escapeinside={(*}{*)},caption={Metameric varifocal phase-only hologram optimisation.},label=lst:Main]
def holographic_metamer((*$\image\target$*), (*$\gaze$*), (*$\distance$*), opt, steps):
    tp = percept((*$\image\target$*),(*$\gaze$*)) # (*See \refLst{Perception}*)

    (*$\phase$*) = define_initial_phase(type='random') 
    (*$\phase$*).requires_grad = True
    
    for i in range(steps):
        optimiser.zero_grad()
        for (*$\lambda$*) in range((*$\lambda_r,\lambda_g,\lambda_b$*)):
            (*$\image\perceived$*) = propagate((*$\phase$*),(*$\wavelength$*),(*$\distance$*)) # (*See \refLst{Display}*)
        pp = percept((*$\image\perceived$*),(*$\gaze$*)) # (*See \refLst{Perception}*)
        loss = l1(pp-tp)
        loss.backward()
        optimiser.step()
        
    return (*\textbf{$\phi$}*)
\end{lstlisting}
\end{minipage}

\noindent\textbf{Main loop:} Our implementation (\refLst{Main}) follows the design of recent hologram optimisation methods~\cite{zhang20173d,chakravarthula2019wirtinger,chakravarthula2021gaze}.
The variable to optimise is a three-channel grid of phase values \phase, initialised randomly.
In an optimisation loop, the holographic image formation is simulated using \texttt{propagate()}, followed by a mapping into a perceptual space \texttt{percept()} (we will discus implementation of both below).
Comparing this percept to the reference percept results in a scalar loss that is back-propagated to the phase values \phase.

All our results were produced using Stochastic Gradient Descent with ADAM \cite{kingma2014adam} as the optimiser on a computer with Intel$^\copyright$ i9 CPU and NVIDIA$^\copyright$ RTX 2080 Ti with 200 steps.
Note that our pipeline can run both on CPU and GPU.
In our case, optimizing a hologram for all colour channels takes 90 seconds in total.
Often, multiple holograms of a similar scene are desired.
In this case, the initialisation of the next rounds of hologram optimisation with a previously calculated hologram can generally decrease this time down to 4 seconds with only five iterations.
The number of iterations required will be somewhat higher if the viewpoint, focus or gaze location change drastically between frames.
Initializing with the previous hologram in this way also improves temporal consistency between holograms, avoiding the flicker that can result from changing between very different metamers at each frame.

\noindent\textbf{Display model:} We model the propagation of light from the \ac{SLM} to the human eye as Fresnel diffraction \cite{born2013principles}, which is typically approximated as a convolution with a non-compact and dense kernel, best implemented using a Fourier transform.
The kernel changes depending on the distance \distance and the wavelength \wavelength.
Hence both are taken into account by \texttt{fresnel\_kernel}.

\begin{minipage}{.45\textwidth}
\begin{lstlisting}[language=iPython,escapeinside={(*}{*)},caption={Propagation from RGB phase to perceived RGB images.}, label=lst:Display]
def propagate((*$\phase$*),(*$\wavelength$*),(*$\distance$*)):
    for (*$\wavelength$*) in (*$\wavelength_\mathrm r, \wavelength_\mathrm g, \wavelength_\mathrm b$*):
        (*$\wave_0$*) = generate_complex_field(1.,(*\textbf{$\phase_\lambda$}*))
        H = fresnel_kernel((*$\distance$*), (*$\wavelength$*))
        (*$\image_{\mathrm p, \wavelength}$*) = ifft(H * fft((*$\wave_0$*)))
    return norm((*$\image\perceived$*), axis = -1)**2
\end{lstlisting}
\label{list:sgd}
\end{minipage}

\noindent\textbf{Perception model:} Our key contribution is a practical, efficient and differentiable mapping, \texttt{percept()}, from an image to a perceptual space that accounts for peripheral perception and is suitable for hologram optimisation (\refLst{Perception}).
This method is called twice, once on the target and in each iteration on the image simulated from the current phase state.
Internally, it uses a function \texttt{pool} for computing local means and variances over regions of varying size using convolution.
This function relies on two components which we briefly describe here. Firstly \texttt{make\_steerable\_pyramid()} recursively and in constant time computes a steerable pyramid following the method described in \cite{simoncelli1995steerable}, in our case with two orientations (vertical and horizontal) using 5$\times$5 kernels.
Secondly, \texttt{make\_lod\_map()} computes a map, that for every pixel holds the level at which a MIP map needs to be read to achieve the correct pooling region size for a specific eccentricity.
The full details of computing this map are given in the supplemental material.

\begin{minipage}{.45\textwidth}
\begin{lstlisting}[language=iPython,escapeinside={(*}{*)},caption={
Mapping an image and a gaze to a perceptual space.
}, label=lst:Perception]
def pool(image,gaze):
    lod_map = make_lod_map(image, gaze)
    mipmap = make_mipmap(image)
    return trilinear_sample(mipmap, lod_map)

def percept((*$\image$*),(*$\gaze$*)):
    p = make_steerable_pyramid((*$\image$*))
    for b in p:
        m = pool(b, (*$\gaze$*))
        s = sqrt(pool(b*b, (*$\gaze$*)) - m*m)
        features.append(m)
        features.append(s)
    return features
\end{lstlisting}
\label{list:loss}
\end{minipage}

\subsection{Holographic display prototype}
\begin{figure}[h!]
\centering
\includegraphics[width=\columnwidth]{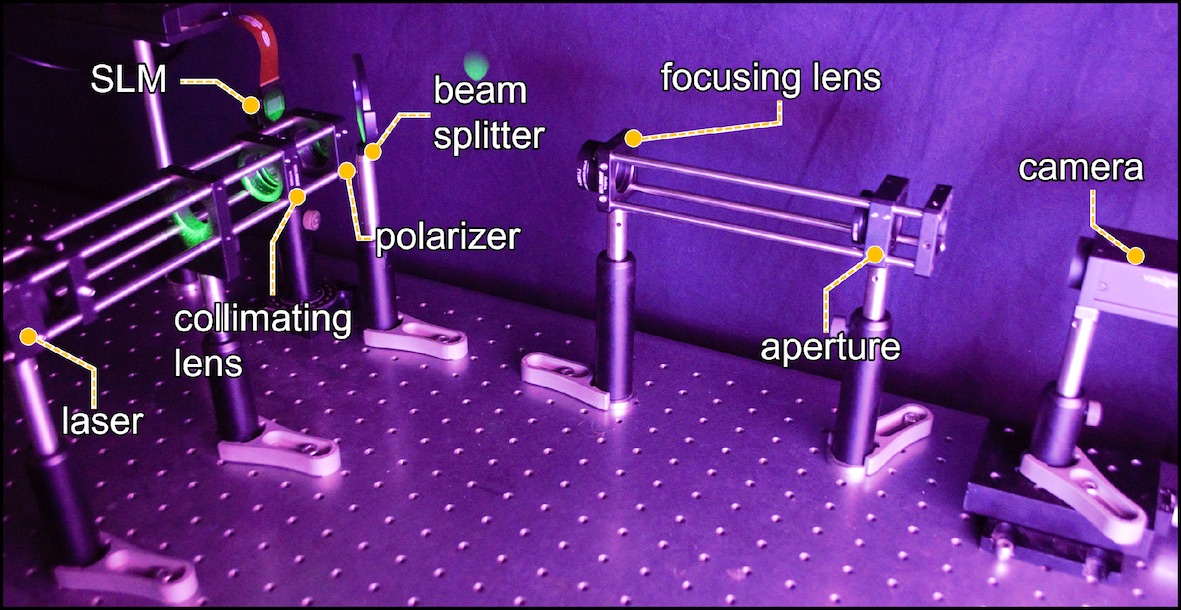}
\caption{
Proof-of-concept holographic display prototype. 
We use a green laser as as a light source. 
We collimate the light from that green laser, and illuminate a phase-only \ac{SLM}. 
The pattern displayed on the \ac{SLM} modulates the phase of the light. 
We observe images reconstructed at various depths using a focusing lens, an aperture and a bare image sensor.
}
\label{fig:display_prototype}
\end{figure}

Having established the theoretical basis of our hologram optimisation pipeline and how to simulate outcomes from this pipeline, we tested using a physical holographic display.
At the time of this manuscript, there was no commodity holographic display that we could purchase off-the-shelf. 
Therefore, we constructed a single colour phase-only holographic display on an optical bench (see \refFig{display_prototype}).

For this purpose, we use a fibre-coupled laser diode (OSI Laser Diode, Inc - TCW RGBS-400R) with an operating wavelength of 520~nm.
We collimate and polarise the laser light source with a Thorlabs LB1945-A bi-convex lens with a 200 mm focal length and Thorlabs LPVISE100-A polariser.
The linearly polarised collimated beam bounces off the beamsplitter, towards our $0.93$ degrees tilted phase-only \ac{SLM}, Holoeye Pluto 2.0 (tilted half order).
To avoid undiffracted light, we add a horizontal grating to the displayed holograms on our \ac{SLM}. 
The horizontally grated hologram, $u_0'$, is
\begin{equation}
  u_0'(x,y) =
  \begin{cases}
            e^{-j(\phase(x,y) + \pi)} & \text{if $\lfloor x\rfloor$ is even} \\
            e^{-j\phase(x,y)}         & \text{if $\lfloor x\rfloor$ is odd} \\
  \end{cases}\;,
  \label{eqn:horizontal_grating}
\end{equation}
where $\phase$, the original phase of $u_0$, is modified. 
This grating ensures that the image reconstructions are not visually affected by the undiffracted beam (0th order reflections).
We capture the reconstructed images from these modulated beams using a Point Grey GS3-U3-23S6M-C USB 3.0 camera and a cascade of beam focusing lenses Thorlabs LA1908-A and LB1056-A. 
We also added a pinhole aperture, Thorlabs SM1D12, in between the camera and these lenses to avoid the undiffracted beam interfering with the result.
In our system, the target image plane for our holograms is about 15cm away from the optical setup. 
We note that our current prototype display is not a complete near-eye display or projection display.
However, it does allow us a way to practically test our method in a safe way, and to verify that the optical reconstructions appear correct on real hardware.

The resolution of holographic displays is only affected by the SLM resolution.
Therefore, the image resolution of the holographic display can be calculated as 8 $\mu$m as lateral and 1 mm as axial spot sizes that are defined by Abbe's law \cite{padmanaban2019stereograms}.   

The full realization of a complete near-eye display would require appropriate eye-piece optics.
A standard eye-piece optic that can be used for such a system can be a lens with a focal length of 50 mm.
This architecture's field of view (FOV) can be calculated as 17.5\textdegree{} horizontal x 9.9\textdegree{} vertical.  

\begin{figure*}
\centering
\includegraphics[width=1.0\linewidth]{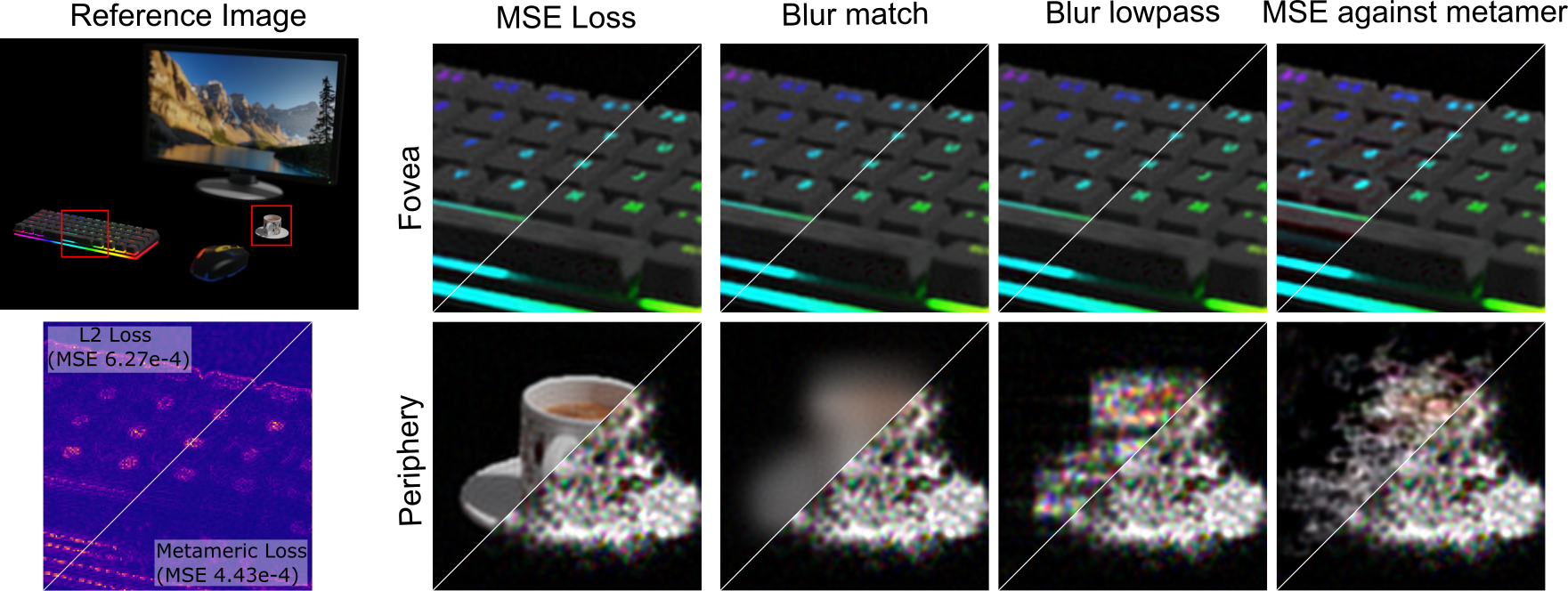}
\caption{
Comparison of different methods in a simulated environment.
Bottom-left compares the output using metameric loss and \ac{MSE} loss to the reference, and gives overall \ac{MSE} losses for this part of the fovea.
Right: output of each method in the fovea and periphery in the upper left triangle,
and our method ($\metamerloss$) in the lower-right triangle for comparison.
The competing methods are: reconstructing images with the correct depth of field (\ac{MSE} Loss), reconstructing using MSE against a blurred image ($\blurmatch$) Blur match, reconstructing using MSE between blurred source and target ($\blurlow$) Blur lowpass, reconstructing a target metamer image (\ac{MSE} against metamer).  
}
\label{fig:ablation}
\end{figure*}

\section{Evaluation}\label{sec:Evaluation}
In this section, we will compare our method to different alternatives in terms of fidelity (\refSec{Comparison}) and demonstrate results on our display prototype (\refSec{Demo}). 

\mysubsection{Comparison}{Comparison}
We compare different methods qualitatively, using the same iteration count on a set of natural images.

\textbf{Methods:} We study four methods that differ only in their loss.
The first is the pixel-value $\mathcal L_2$ loss in RGB.
The second and third are na\"ive baseline foveated losses inspired by \cite{surace2021learning}, which account only for the acuity of the visual system in the periphery.
We study two variants.
The first loss $\blurmatch$ convolves the target $T$ according to the acuity-dependent blur kernel $B$ and compares this to the reference $I$ using $\mathcal L_2$:
\begin{equation}
 \blurmatch := \mathcal{L}_2(I, B(T)) 
\end{equation}

The second $\blurlow$ blurs the result of the optimisation accordingly and compares this to the target.
\begin{equation}
    \blurlow := \mathcal{L}_2(B(I), B(T))
\end{equation}
We note that although the definitions of these two losses are very similar, they behave in very different ways.
The $\blurmatch$ loss enforces the image $I$ to exactly match a blurred target $B(T)$ and have no high frequency content in the periphery.
In contrast $\blurlow$ only enforces the low frequency content of $I$ in the periphery to match $B(T)$, and does not constrain the higher frequencies of $I$ away from the fovea.

The fourth method, $\metamerl$, is $\mathcal{L}_2$ loss against a metamer generated using the method of \cite{walton2021beyond}. 
We note this is not the same as metameric loss. 
Our metameric loss only constrains the output to be any metamer of the target. 
This loss constrains the output to be identical to one particular metamer of the target.

The final method is our metameric loss $\metamerloss$ (see eqn.~\ref{eqn:metameric_loss}).

\textbf{Data:} We study results on our dataset, which consists of both natural images and rendered virtual scenes.
This was produced as outlined in \refSec{Implementation}.

\textbf{Metric:} As there is no reliable metric to capture the perceptual effects of focus and fixation, the evaluation has to rely on qualitative examples.
To judge the quality, we show insets from the fovea as well as insets from the periphery.
All comparisons are made after the same number of gradient descent steps (200).
We note that the foveated results from the metameric and blur losses are intended to be viewed whilst fixated on the intended gaze location.
As such, whilst the foveal insets can be compared directly to the reference to assess quality, the peripheral insets should be compared by fixating at a different location and keeping them in the periphery of one's vision.

\textbf{Results:} Results of all five methods on a natural scene are shown in \refFig{ablation}.

The standard $\mathcal{L}_2$ loss distributes error uniformly across the image.
This approach would likely be the best if the user's gaze were not known.
However, if gaze information is available, it does not prioritise the fovea in any way, meaning visible artefacts will still be uniformly distributed and presented there in the case of an actual holographic display. In this case, the fovea is noticeably blurry and some noise can be seen.

$\blurmatch$ produces images with a similar quality in the fovea, although the periphery is naturally very blurry.
As noted in \cite{walton2021beyond} this is acceptable for low levels of blur. 
However, as blur increases, the lack of high frequencies becomes increasingly noticeable, even when fixating in the correct location.
As with $\mathcal{L}_2$ it does not prioritise the fovea.

The $\blurlow$ also has similar quality to $\mathcal{L}_2$ in the fovea.
However, this loss does not restrict the higher frequency content in the periphery.
As a result, the appearance of the periphery can vary greatly depending on the optimisation task.
This application produced disturbing grid-like noise that we found to be visible even when in the periphery of vision.

The $\metamerl$ loss approximated a metamer of the target image.
However, it in no way prioritises the foveal region.
Consequently, the result will not necessarily appear superior to standard $\mathcal{L}_2$ loss against the original target image, even when fixating in the correct location. 
In fact in this case the fovea appears noticeably worse than the previous approaches, possibly because the metamer generated using the approach of \cite{walton2021beyond} is harder to approximate with a holographic reconstruction than the original target image.

The metameric loss $\metamerloss$, like $\blurlow$ does not enforce pixel-level accuracy to the target image.
Unlike $\blurlow$ however, it requires that the orientation statistics in the periphery match the target.
This extra constraint results in a periphery that appears similar to the metamers of \cite{walton2021beyond}.
The metameric loss tolerates some degree of noise, ringing or other artefacts in the periphery, making it more flexible than $\mathcal{L}_2$ whilst still forcing the output to be close to the target perceptually.
In this case, the extra flexibility and the fact the loss prioritises the fovea have allowed it to produce a sharper result with fewer artefacts in the fovea, and perceptually correct content elsewhere.

\begin{figure*}[t]
\centering
\includegraphics[width=1.0\linewidth]{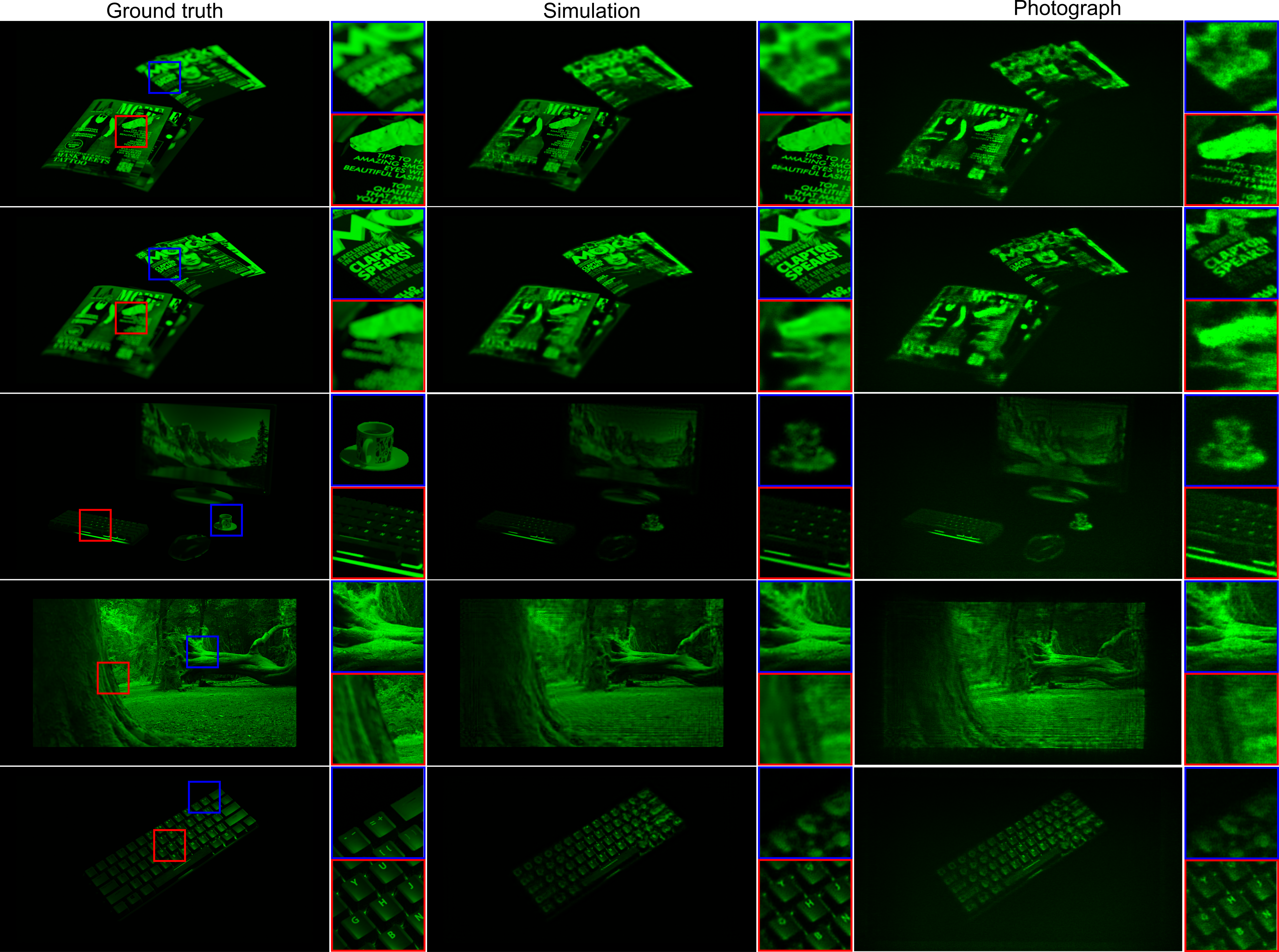}
\caption{
Metameric varifocal holographic reconstructions for holographic displays.
The first column provides ground truth images that are going to be metamerised.
The second column shows simulated image reconstructions of optimised holograms derived using our metameric varifocal hologram optimisation pipeline.
The last column shows captured photographs of image reconstructions of our holograms when displayed in our proof-of-concept display prototype.
For all columns, insets that shows zoomed-in versions of foveal and peripheral regions are also provided.
}
\label{fig:prototype_results}
\end{figure*}

\mysubsection{Prototype results}{Demo}
Now that we have established our metameric loss (not MSE against a metamer) and hologram optimisation methods, we assess the outcome of our entire pipeline in an actual holographic display.  

\begin{figure}
    \centering
    \includegraphics[width=\columnwidth]{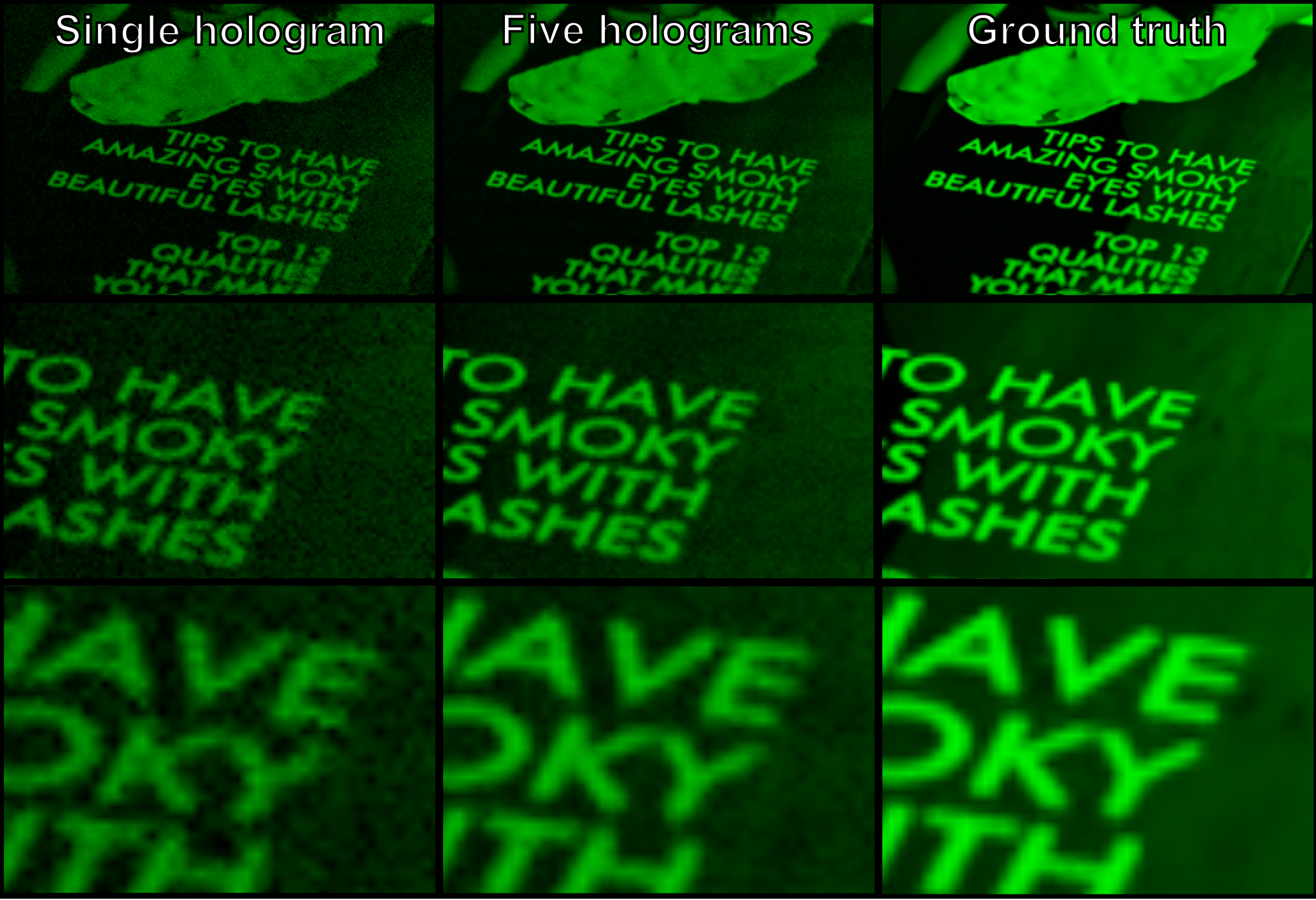}
    \caption{
    Simulated Temporal averaging.
    Rather than using a single hologram based image reconstruction (images on the  left column), we rely on showing multiple images (center column).
    This way, we can generate images that suffer less from noise. 
    Note that we use five holograms for this temporal averaging example.
    }
    \label{fig:results-averaging}
\end{figure}

A temporal averaging approach is commonly used in holographic displays to avoid speckle noise and improve the results' perceived quality when showing static images~\cite{curtis2021dcgh}.
Rather than displaying a single, static hologram to a user, holograms optimised with slightly modified target images are rapidly displayed in sequence.
These images have different realisations of the high-frequency noise and, if displayed at a rate above the user's \ac{CFF} threshold, will be ``averaged" by the \ac{HVS}.

There is one issue to address when using this averaging approach with the metameric loss.
We note that in general, the mean of two distinct metamers to an image is not a metamer.
Taking the mean of multiple, very different metamers to an image tends to produce a blurry result in the periphery.
For this reason, we need to ensure that we use the same metamer for a particular target image in our temporal averaging approach, just adding different frequencies of high-frequency holographic noise in each case.
We achieve this by initializing with a noise phase image to create the first hologram, and for the subsequent holograms we initialise with the previous optimised phase, optimising for just 5 iterations with a lower learning rate.
This has the added benefit of reducing the number of iterations needed to generate the subsequent holograms.

\refFig{results-averaging} shows an example.
Here we show a single hologram and the average of five consecutive holograms with a different noise, simulating the image perceived by the user.
Our current prototype display operates at 60Hz, meaning that with 5-frame temporal averaging the effective framerate is reduced to 12Hz.
Note, however, that the noise is greatly reduced and the image closer to the ground truth on the right.
One of the goals in \ac{CGH} is to avoid using multiple holograms so that holographic displays can use the total frame rate of an \ac{SLM}.
Our metameric loss improves the precision of a single hologram, as later shown in \refFig{prototype_results}, helping towards meeting the goals of \ac{CGH}.

We provide sample image reconstructions of our holograms optimised using our metameric varifocal hologram generation pipeline as in \refFig{prototype_results}.
These samples contain a set of ground truth images that are targeted to be metamerised.
Gaze locations and focal planes ($\sim 0.15m$) are also provided for each sample during an optimisation session.
We intentionally choose images varying from synthetic sparsely populated cases to densely populated natural photographs.
In the figure, the results of our simulated image reconstructions are provided for the holograms optimised using our pipeline. 
Finally, we also provide photographs of image reconstructions from our proof-of-concept display.
Though it is not pixel-perfect, note that simulated cases and actual photographs very closely resemble each other, and the noise patterns at both images are close in terms of spatial distribution.
How well the actual photographic results match the simulation is heavily dependent on the model used in optimisation.
We relied on a state-of-the-art model from the literature to achieve the best results possible at the present time.
As models continue to improve, our pipeline can take advantage of improvements to combat noise in \ac{CGH} in the future.

Focal planes used in our captures varied between 0.14m to 0.16m.
These distances are suitable for virtual reality and augmented reality applications. 
For example, in a typical virtual reality display, a magnifier glass typically of focal length from 35 to 60mm is used between a flat image modulator and the eye.
An image volume in the order of a few millimetres translates to covering virtual image distances from very close distances (10cm) to far away (6m and beyond).
Thus, our \ac{CGH} pipeline can present images at a wide range of focal distances.

\section{Discussion}
Both the simulations and the results from our physical prototype further our understanding of the technical challenges in deriving a metameric varifocal hologram pipeline.
These efforts also lay the foundation of future holographic displays that are perceptually guided.
Nevertheless, more work remains before we can achieve fully practical \ac{CGH}.

\textbf{Interactive rates:} Our current \ac{CGH} optimisation does not run at interactive rates, as is typical for most \ac{CGH} optimisation pipelines.
In the meantime, there has been a push towards taking advantage of learning approaches in \ac{CGH} in recent years~\cite{peng2020neural,shi2021towards,chakravarthula2020learned}.
While learning-based acceleration is outside of the scope of our work, our metameric loss appears eminently suited for learning frameworks.

\textbf{Metameric loss:} Our readers may argue that setting a metamer or a full resolution image as a target image and optimizing a hologram using \ac{MSE} loss can lead to a faster optimisation routine. 
This method is indeed computationally efficient but, as shown in \refSec{Evaluation} above, produces results with similar or worse quality than regular $\mathcal{L}_2$ loss in the fovea.
Also, in an actual or simulated holographic display, perfectly matching target images may be physically impossible, and visual imperfections such as noise are very likely to occur, as can be observed in \refFig{results-averaging}.
Rather than using \ac{MSE} loss against a metamer target or a complete resolution target, optimizing holograms with our metameric loss guarantees metamer solutions that play well with a given holographic light transport model in a simulation.
This advantage clearly shows in our simulated results; however, at the time being, the advantages of our method and other foveation methods for \ac{CGH}~\cite{chakravarthula2021gaze} do not always translate to the real world due to the lack of accurate-enough holographic light transport and display models in the literature -- which is beyond the scope of our focus of this work.
We argue that the importance of metameric loss will become evident as the holographic light transport models match physical hardware accurately in the future.

\textbf{Varifocal visuals:} There has been a long-standing debate about the qualities of a display when it comes to supporting optical depth cues. 
A recent survey on displays~\cite{koulieris2019near} captures a detailed background of this debate.
The varifocal approach in displays has recently proven to improve visual comfort~\cite{johnson2016dynamic}.
Fortunately, the latency requirements for a varifocal display are not demanding as the accommodation duration of human eyes  has been measured as between 500~ms and 1~s in various studies~\cite{campbell1960dynamics,bharadwaj2005acceleration}.
One disadvantage of varifocal displays is that they cannot handle rare cases where gaze does not uniquely determine the focal depth.
Nevertheless, we argue that varifocal displays are strong candidates for supporting optical depth cues in displays.

\textbf{Gaze-contingent displays:} A complete gaze-contingent display requires eye-gaze tracking systems to be involved in the process of generating visuals. 
The accuracy requirements of a gaze-contingent varifocal display have recently been systematically studied~\cite{dunn2019required}.
Unfortunately, our proof-of-concept prototype is not equipped with eye-gaze tracking hardware~\cite{li2020optical,angelopoulos2020event}.
Our work assumes that such eye-gaze information is readily available. 
We calculate holograms with this assumption. 

Real-world gaze trackers suffer from varying degrees of inaccuracy and latency. 
Our method could be adapted to better tolerate these issues, by modifying the foveation and increasing the size of the foveal region.
There is however a trade-off between foveal size and improvement in image quality over MSE loss.

Though some hurdles remain in our implementation, our work resembles a reliable blueprint for a practical \ac{CGH} pipeline that can deliver perceptually accurate visuals to users.


\section{Conclusion}
The versatility in generating high-resolution and three-dimensional visuals makes Computer-Generated Holography a powerful technique suitable for next-generation displays.
However, among many technical issues, achieving three-dimensional visuals with \ac{CGH} still poses a significant challenge in real holographic displays.
We argue that gaze-contingent \ac{CGH} can be key to achieving practical holographic displays with perceptually accurate three-dimensional visuals.
For this purpose, we build upon state-of-the-art perceptual graphics.
We formulate a new differentiable hologram optimisation pipeline with a perceptually guided loss function.
Rather than reconstructing imperfect three-dimensional scenes, our \ac{CGH} method reconstructs visuals at the user's focus.
It offers improved image quality at the fovea, while displaying true metamers of target images in the periphery.
Using gaze-contingency, we formulate our phase optimisation as a two-dimensional problem, removing the need to match a light field or multiplane image.
In this way, our method paves the way towards a practical display that provides perceptually accurate three-dimensional visuals more efficiently.

\acknowledgments{
The authors thank the reviewers for their useful feedback.
The authors also thank Duygu Ceylan for the fruitful and inspiring discussions improving the outcome of this research, and Selim Ölçer for helping with the fibre alignment of laser light source in the proof-of-concept display prototype.
This work was partially funded by the EPSRC/UKRI project EP/T01346X/1 and Royal Society's RGS\textbackslash R2\textbackslash 212229 - Research Grants 2021 Round 2.
}

\bibliographystyle{abbrv-doi}

\bibliography{paper}

\end{document}